\documentclass[a4paper,11pt]{article}
\usepackage{graphicx}
\usepackage{a4wide}
\usepackage{authblk}
\usepackage{mathptmx}      
\newcommand{\Hbar}{\mbox{$\overline{\mathrm H}$}}
\newcommand{\pbar}{\mbox{$\overline{\mathrm p}$}}
\newcommand{\nuHFS}{$\nu_{\mathrm{HFS}}(\overline{\mathrm H})$}
\begin{document}

\title{Measurement of the hyperfine structure of antihydrogen in a beam\thanks{Supported by the European Research Council grant no. 291242-HBAR-HFS, the Austrian Ministry for Science and Research, and the Grant-in-Aid for Specially Promoted Research (19002004), JSPS.}
}


\author[1]{E. Widmann \thanks{ew@antihydrogen.at}}
\author[1]{M. Diermaier}
\author[1]{B.~Juhasz}
\author[1]{C.~Malbrunot}
\author[1]{O. Massiczek }
\author[1]{C.~Sauerzopf}
\author[1]{K. Suzuki}
\author[1]{B.~W{\"u}nschek}
\author[1]{J. Zmeskal}
\author[1,2]{S.~Federmann} 
\author[3]{N. Kuroda}
\author[4]{S. Ulmer}
\author[4]{Y.~Yamazaki} 	


\affil[1]{\it Stefan Meyer Institute for Subatomic Physics,
              Austrian Academy of Sciences,
              Boltzmanngasse 3, 1090 Vienna, Austria
              }

\affil[2]{		\it			CERN, 1211 Geneva 23, Switzerland}
\affil[3]{		\it		Institute of Physics, University of Tokyo, 3-8-1 Komaba, Meguro-ku, Tokyo 153-8902, Japan}
\affil[4]{	\it	RIKEN Advanced Science Institute, Hirosawa, Wako, Saitama 351-0198, Japan}

\date{\today}

\maketitle

\begin{abstract}
A measurement of the hyperfine structure of antihydrogen promises one of the best tests of CPT symmetry. We describe an experiment planned at the Antiproton Decelerator of CERN to measure this quantity in a beam of slow antihydrogen atoms.

\begin{description}
\item[keywords] Antihydrogen; Precision spectroscopy; CPT 
\item[PACS] 36.10.-k, 11.30.Er, 32.10.Fn
\end{description}

\end{abstract}

\section{Introduction}

Antihydrogen is the simplest atom consisting entirely of antimatter. Since its hydrogen counterpart is one of the most precisely measured atoms in physics, a comparison of antihydrogen and hydrogen offers one of the most sensitive tests of CPT symmetry. This project proposes to measure the ground state hyperfine splitting \nuHFS\ of antihydrogen (\Hbar), which is known in hydrogen to be 
$\nu_{\mathrm{HFS}}(\mathrm H) = 1~420~405~751.768(1)$ Hz 
(relative precision $7\times10^{-13}$) as determined in a hydrogen maser \cite{Karshenboim:02,Ramsey:QED} . It is the central part of the program of the ASACUSA collaboration at CERN-AD. Since the maser confines atoms in a teflon coated box which is currently not feasible for antimatter, the experimental method consists of the formation of an antihydrogen beam and a measurement using a spin-flip cavity and a sextupole magnet \cite{Widmann:2001fk,Widmann:2004zr} as spin analyzer like it was done initially for hydrogen. A major milestone was achieved in 2010 when antihydrogen was first synthesized by ASACUSA in a so-called cusp trap \cite{Enomoto:2010uq}. In the first phase of this proposal, an antihydrogen beam will be produced and the \Hbar\ hyperfine splitting will be measured to a precision of below $10^{-6}$ using a single microwave cavity. 

In a second phase, the Ramsey method of separated oscillatory fields will be used to increase the precision by about one order of magnitude. In parallel methods will be developed towards trapping and laser cooling the antihydrogen atoms. Letting the cooled antihydrogen escape in a field free region and performing microwave spectroscopy offers the ultimate precision achievable to measure \nuHFS\ and one of the most sensitive tests of CPT. Assuming the antihydrogen atoms would pass a height difference of 1 m in an atomic fountain, a line width of about 1 Hz will be feasible corresponding to a relative accuracy of $10^{-9}$. The time needed to achieve this goal, however, exceeds the scope of the current proposal.

\section{Antihydrogen and CPT symmetry}
\label{sec:1}

Fundamental symmetries and resulting conservation laws are a fundamental concept in physics. In particle physics, it was believed that Nature conserves the symmetries of space until the discovery of parity (P) and later charge conjugation and parity (CP) violation. One of the cornerstones of the standard model (SM) of particle physics is the combined application of charge conjugation, parity, and time reversal: CPT. The proof that this symmetry is conserved in the SM originates from the properties of the quantum field theories used and is based on a mathematical theorem. Whether the conditions generating this mathematical theorem are valid in Nature is an important question to answer experimentally. 

\begin{figure}[t]
\begin{center}
  \includegraphics[width=.7\textwidth]{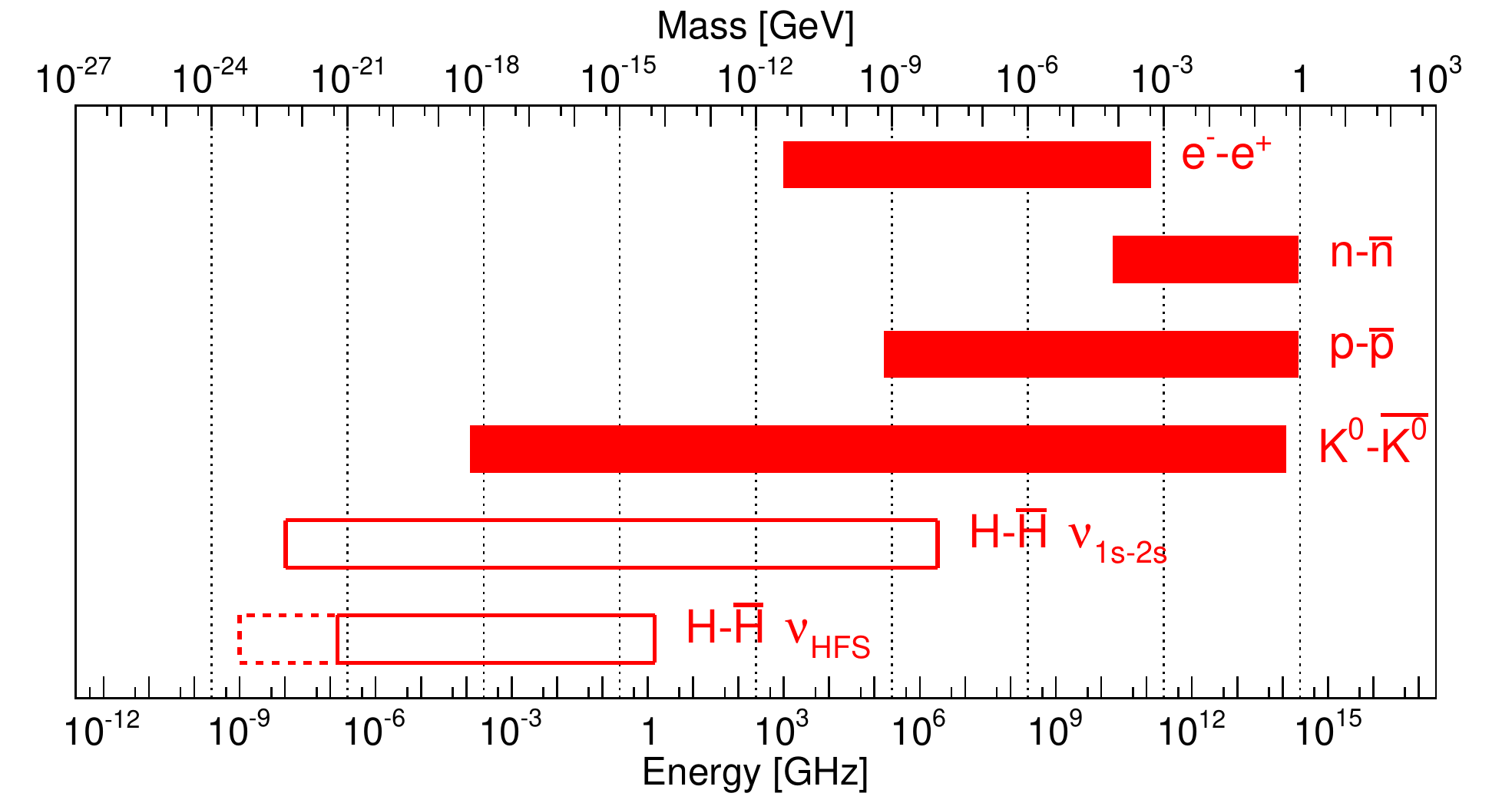}
\caption{Comparison of certain CPT tests. The right end of the bar gives the absolute value of the measured quantity, its length the relative precision of the measurement, and the left end the sensitivity on an absolute scale. Compared are the masses of several elementary particles and the two transitions in antihydrogen, provided they could be measured to the same precision as hydrogen. For the HFS the solid line shows the limits reachable in the first phase of an atomic beam, the dashed line the one achievable in an atomic fountain with 1 m height difference.}
\label{fig:CPT1}       
\end{center}
\end{figure}

Further interest arises in studies of CPT symmetry from the fact that certain prerequisites of the mathematical proof like point-like particles are not any more valid in string theory, so that an observation of CPT-violation could be a first hint for the validity of string theory and a clear evidence for physics beyond the standard model.
Based on an argument that any CPT-violating interaction must manifest itself by a term in the Dirac equation that has the dimension of mass or energy, the highest sensitivity can be obtained by reaching the highest precision in terms of absolute energy. This argument has been brought forward by Kostelecky et al. in their extension of the standard model SME \cite{Colladay:1997vn,Bluhm:1999vq}, which is the only model that allows a comparison of CPT tests in different systems. Often the mass difference between the neutral kaon and antikaon is referred to as the best test of CPT since it is known with a relative precision of $6\times10^{-19}$ \cite{Beringer:2012ys}. On an absolute mass scale, this value corresponds to a frequency of 100 kHz, which is much larger than what can be achieved in high-precision spectroscopic measurement of atomic transitions. Among them, the antihydrogen hyperfine transition could potentially obtain one of the best values on an absolute scale (see Fig.~\ref{fig:CPT1}).

Like CP violation, which is observed so far only in K and B meson decays, CPT violation may also appear only in a certain area of particle physics. Therefore measurements involving all types of particles and interactions (in Fig.~\ref{fig:CPT1} results are listed for leptons, baryons, meson, and atoms) are needed.
The study of the \Hbar\ hyperfine structure is complementary to the 1S-2S laser spectroscopy. While the latter is dominated by the electromagnetic interaction and the energy levels are in first order sensitive to the well known positron mass (which dominates the reduced mass), the HFS is caused by the spin-spin interaction. From a relative precision of $\sim 10^{-5}$, correction terms arising from the finite antiproton (\pbar) charge and magnetic radius play a role and the comparison between hydrogen and antihydrogen is then also sensitive to possible strong-interaction induced differences in the internal structure of proton and antiproton.


\section{Experimental method}
\label{sec:2}

\begin{figure}[b]
\begin{center}
  \includegraphics[width=8cm]{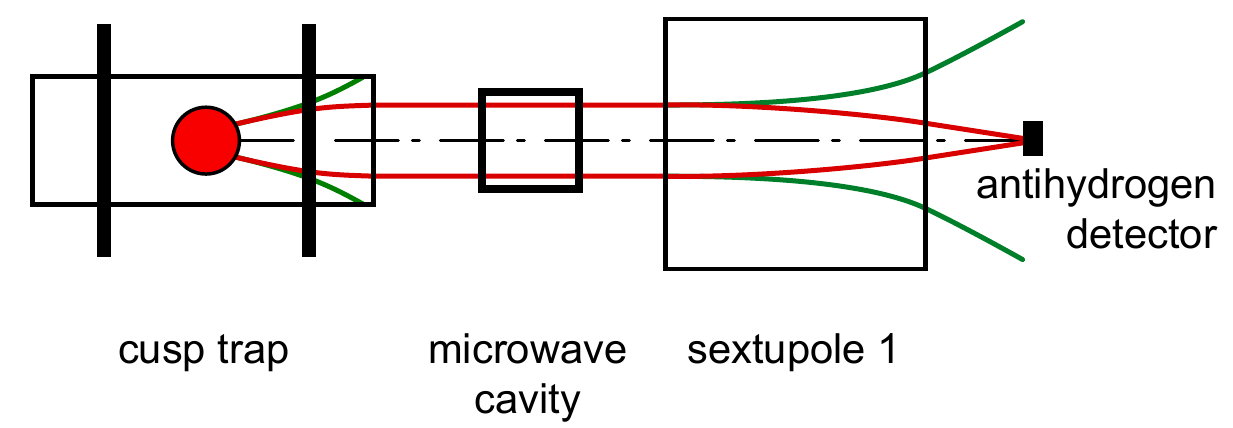}
\caption{Schematic setup of the atomic beam line used in the experiment.}
\label{fig:beamline1}       
\end{center}
\end{figure}

The basic approach is in the first stage to follow the methods used in the early days of hydrogen by Rabi and others \cite{Rabi:34a} to use an atomic beam. It has the advantage that an antihydrogen beam of temperature up to 100 K can be used, while the trapping of neutral antihydrogen atoms requires temperatures $\le 0.5$ K. The drawback is that in a beam experiment the achievable resolution is determined by the flight time of the antihydrogen atoms through the cavity (see Fig.~\ref{fig:beamline1}). The best value obtained for hydrogen in an atomic beam is $4 \times 10^{-8}$ \cite{Prodel:52}.

In order to produce an antihydrogen beam, members of ASACUSA proposed to use a type of trap called Òcusp trapÓ \cite{MohriEP} which uses an inhomogeneous magnetic field created by two Helmholtz coils with the currents running in opposite direction to confine charged and neutral particles. A major breakthrough has been achieved in 2010 when for the first time antihydrogen production in the cusp trap \cite{Enomoto:2010uq} could be demonstrated, enabling now a start of the actual measurement campaign. Due to the inhomogeneous magnetic field of the cusp trap, the \Hbar\ beam emerging is expected to have a polarization of $\sim 30$ \% \cite{Enomoto:2010uq}.

The experiment will be performed using the same principle methods as previous ones on hydrogen, but adapted to the experimental conditions of antihydrogen. A major difference is the scarcity of antihydrogen atoms, leading to the necessity of designing atomic beam lines with much larger transmission than used before. This requires cavities and sextupole magnets with apertures of 10 cm, which have already been constructed. On the other hand antimatter can be detected with much higher efficiency due to its characteristic annihilation pattern. The different parts of the experimental setup will be described in the following.

%
%
%



%

\subsection{Atomic beam line}
\label{sec:2.2}

\begin{figure}[hb]
\begin{center}
  \includegraphics[width=.8\textwidth]{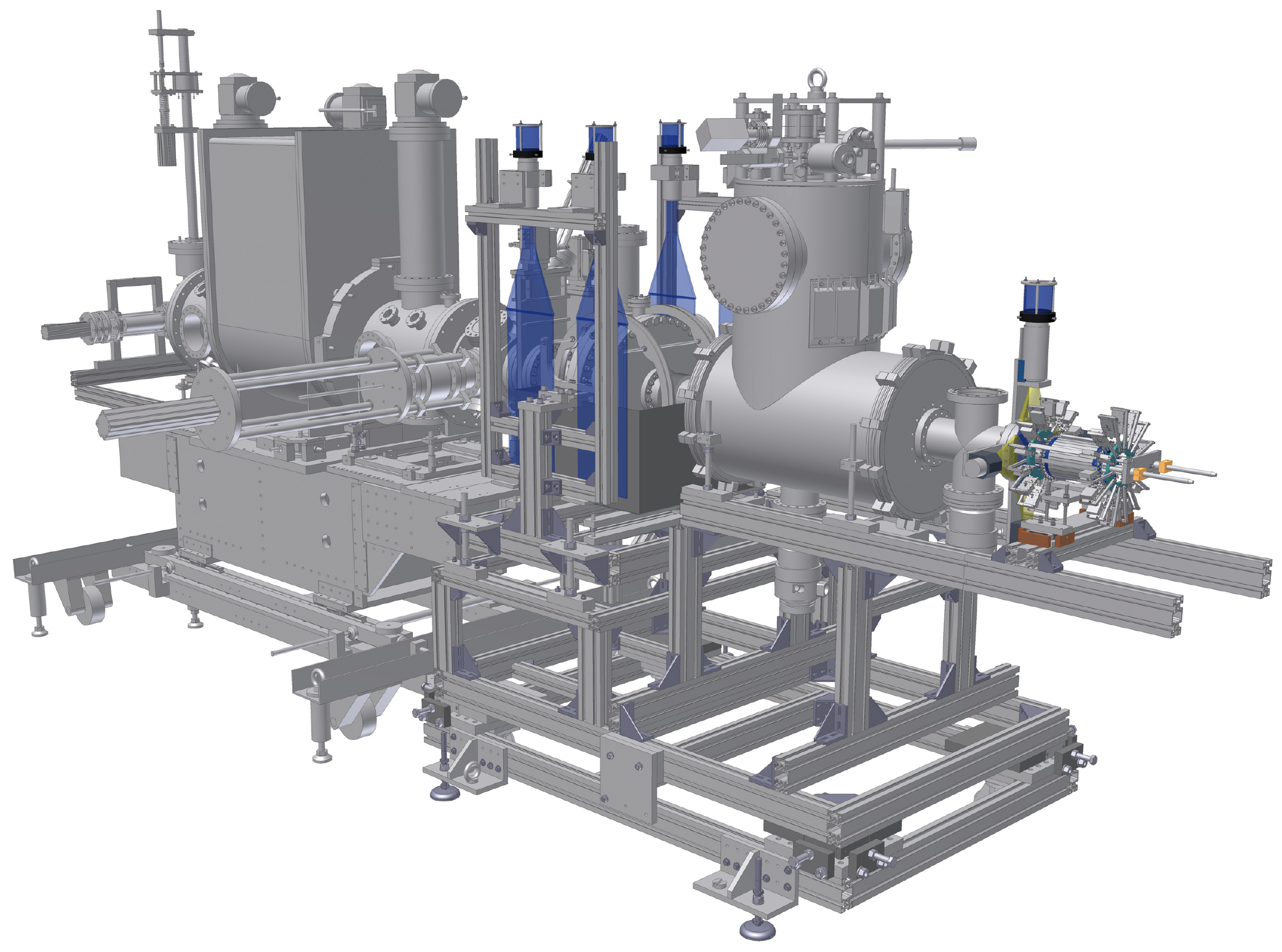}
\caption{View of the set up with the cusp trap to the left and the sextupole and antihdyrogen detector to the right. The cavity is surrounded by plastic scintillation counters (blue) for normalization purposes.}
\label{fig:beam line}       
\end{center}
\end{figure}

Fig.~\ref{fig:beam line} gives a view of the atomic beam line including the cusp trap. The main components are a microwave cavity and a superconducting sextupole magnet, both with an opening of 10 cm diameter for the \Hbar\ beam. In order to generate a microwave field oscillating at 1.42 GHz with a homogeneity of better than 10\% over a cylindrical volume of 10 cm diameter and 10 cm length at a wavelength of 21 cm, a strip line cavity was chosen \cite{Juhasz:2009dh,Federmann:2012ly} where the faces of the cylinder are covered by meshes to let the \Hbar\ atoms pass but contain the microwaves. The cavity was built and successfully operated.

The width of the resonance line is given by the time of flight of the atoms passing through the cavity. Assuming a \Hbar\ temperature of 50 K, the most probable velocity being 1000 m/s, the atoms take about $T=100$ $\mu$s leading to a line width of $\Delta f \sim 1/T = 10$ kHz. This corresponds to a relative width of $\Delta f/f = 10$ kHz/$1.42$ GHz = $7\times 10^{-6}$. Provided the line center can be determined to an order of magnitude better than the width if enough statistics is available, a relative precision of better than $10^{-6}$ is feasible.

\begin{figure}[t]
\begin{center}
	\includegraphics[width=8cm]{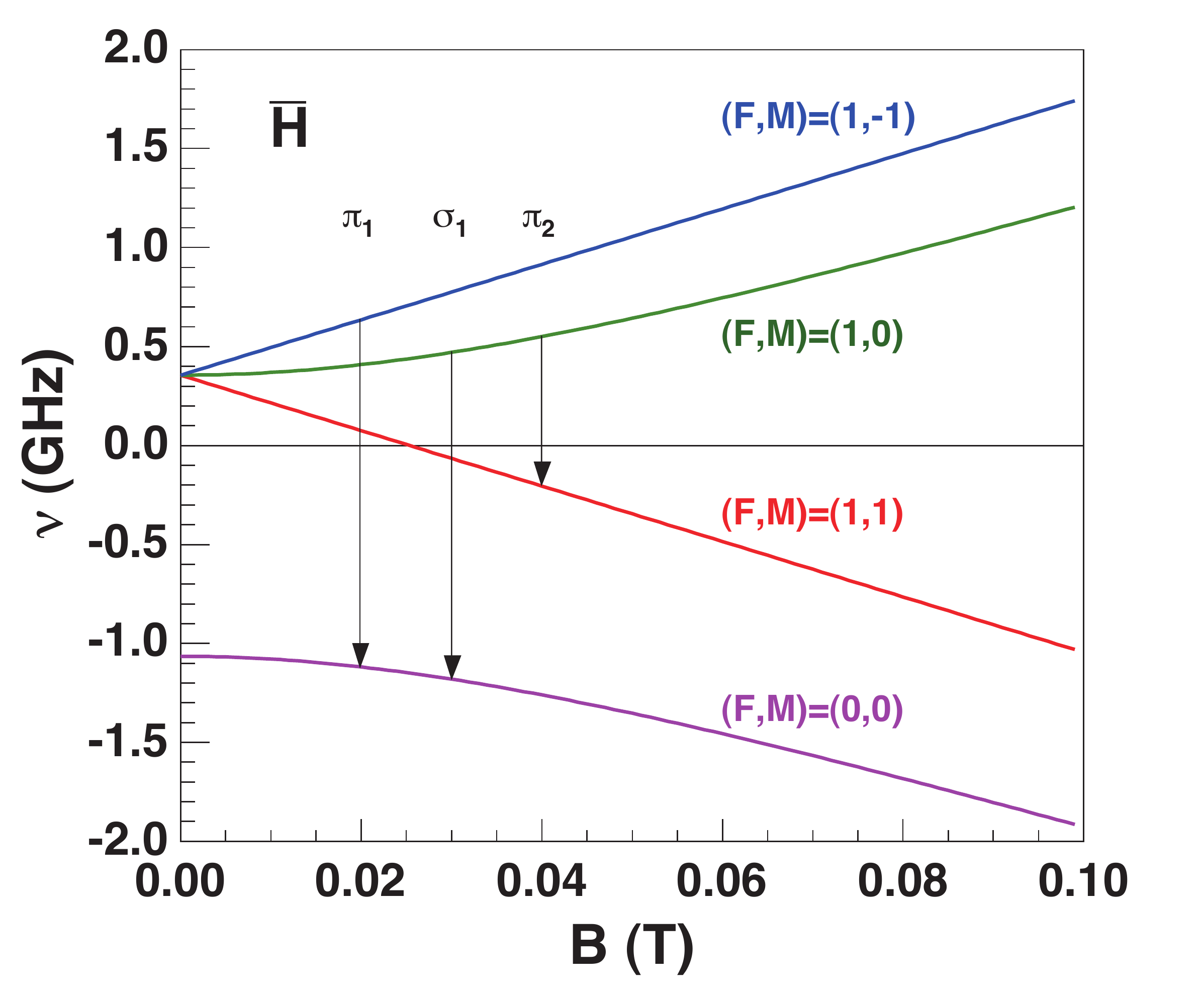}
\caption{Breit-Rabi diagram for antihydrogen showing the three possible transitions from low-field to high-field seeking states.}
\label{fig:BR}       
\end{center}
\end{figure}

Like in the standard beam experiments with hydrogen, a small constant magnetic field of the order $B=1$ Gauss is necessary to avoid Majorana spin flips, leading to a splitting of the hyperfine levels as shown in Fig.~\ref{fig:BR}. It is provided by Helmholtz coils surrounding the cavity which are once more surrounded by two layers of magnetic shielding (mu-metal, thickness 1 mm) to suppress the stray magnetic fields of the other magnets in the area as well as the earth magnetic field. Using this configuration, static magnetic fields of homogeneity of $\sim 5$\% could be achieved \cite{Federmann:2012ly}.

\begin{figure}[t]
\begin{center}
  \includegraphics[width=\textwidth]{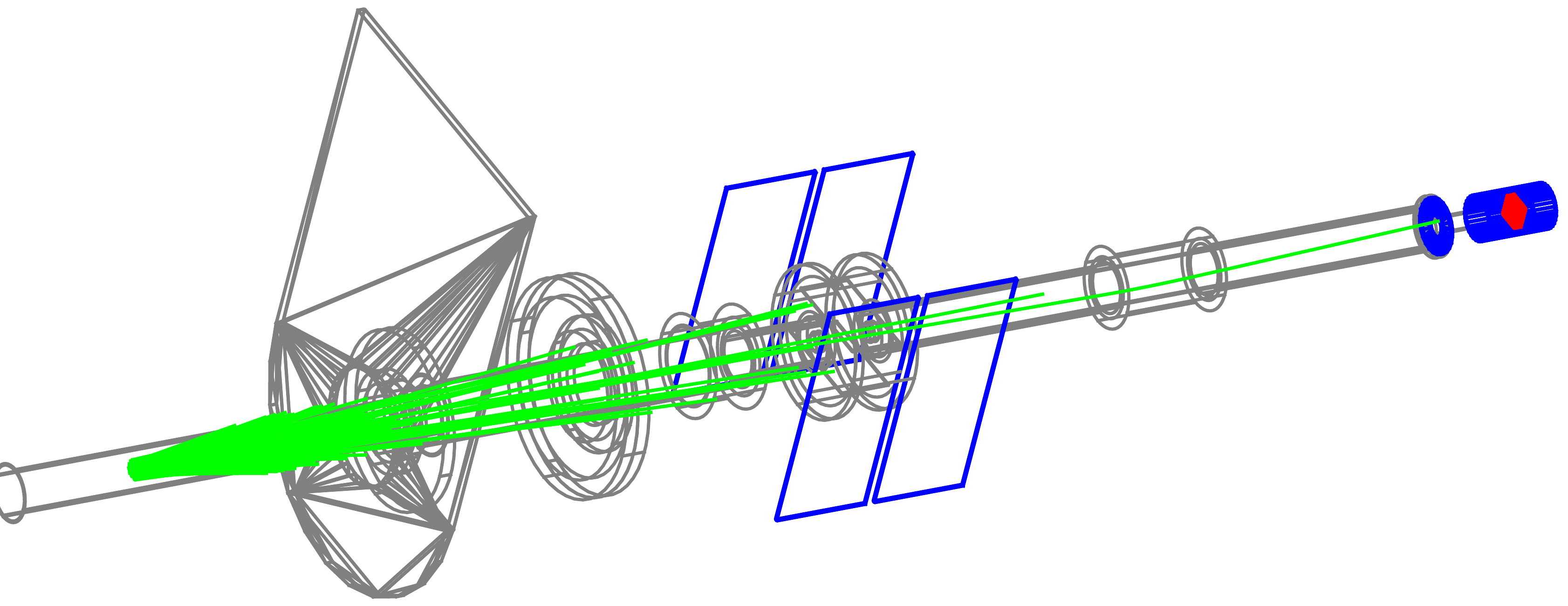}
\caption{Trajectories of \Hbar\ atoms (green) in the beam line emulated by a modified version of Geant4 which is able to track neutral atoms with a magnetic moment through field gradients. Parts of the cusp trap, the cavity, and sextupole are shown in gray, while the blue color denotes detectors for annihilation products. The red dot is the central antihydrogen detector (see sec.~\ref{sec:2.2}).}
\label{fig:geant4}       
\end{center}
\end{figure}

For spin selection and focussing, a superconducting sextupole with 10 cm bore, a maximum pole field of 3.5 T and effective field length of 22 cm was purchased from Tesla Engineering Ltd. (UK). The sextupole will focus low-field seeking \Hbar\ atoms and defocus high-field seeking ones.

The trajectories of \Hbar\ atoms were simulated assuming isotropic direction and a velocity distribution following a Maxwell-Boltzmann shape at the formation point (cf. Fig.~\ref{fig:geant4}). The simulation showed that about $10^{-4}$ of all low-field seekers produced arrive at the antihydrogen detector. At a typical antihydrogen production rate of 100 \Hbar /s, the count rate at the antihydrogen detector will be one event in 1--2 minutes. Thus the antihydrogen detector must be able to distinguish \Hbar\ events from cosmic background and pions from antiproton annihilations in other parts of the setup.

\subsection{Particle detectors}
\label{sec:2.2}

We use plastic scintillation detectors for charged particles. There are three types of them: 

\paragraph{Normalisation counters} are used to measure the flux of antihydrogen atoms in the beam line. They are surrounding the cavity, where the center of the meshes are solid over a diameter of 4 cm to block \Hbar\ atoms that are too close to the centre of the beam pipe. Since the field in the center of the sextupole is zero and rises quadratically, the bending power in the center is too small to separate high field seekers and low-field seekers and atoms along this path would therefore produce background. The number of \Hbar\ atoms annihilating at the beam blockers are measured during the experiment providing a signal proportional to the antihydrogen flux. 

\paragraph{Veto counters} are placed downstream of the sextupole magnet to detect the high-field seeker atoms deflected towards the outside by the sextupole magnet. The counters surround a circle of 4 cm diameter, letting predominantly low-field seeking atoms pass towards the antihydrogen detector. 

\paragraph{The antihydrogen detector} needs to distinguish \Hbar\ annihilations from pions and cosmic muons and electrons. In order to achieve almost unity detection efficiency, only the annihilation of \pbar\ into charged particles (in average three pions) are detected. At the end of the beam pipe the antihydrogen atoms hit a steel plate of 4 cm diameter and 2 mm thickness and annihilate. Immediately behind the plate 64 small blocks of plastic scintillator ($6\times6\times20$ mm$^{3}$) are placed which are connected to a 64 channel PMT (Hamamatsu H8500C). A hit in one of these detectors is used as master trigger for the data acquisition. 

\begin{figure}[t]
\begin{center}
  \includegraphics[width=.5\textwidth]{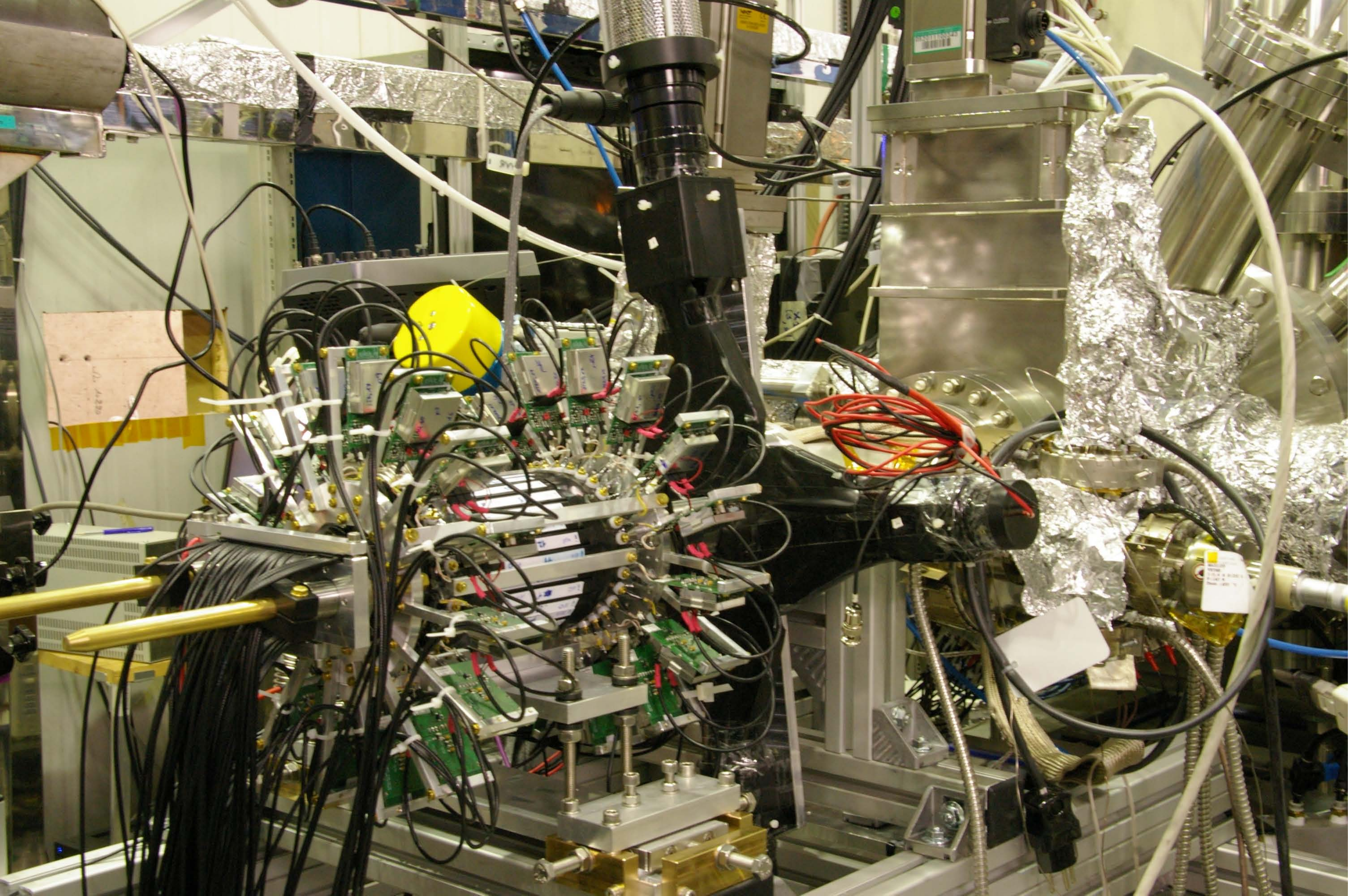}
    \includegraphics[width=.45\textwidth,height=.33\textwidth]{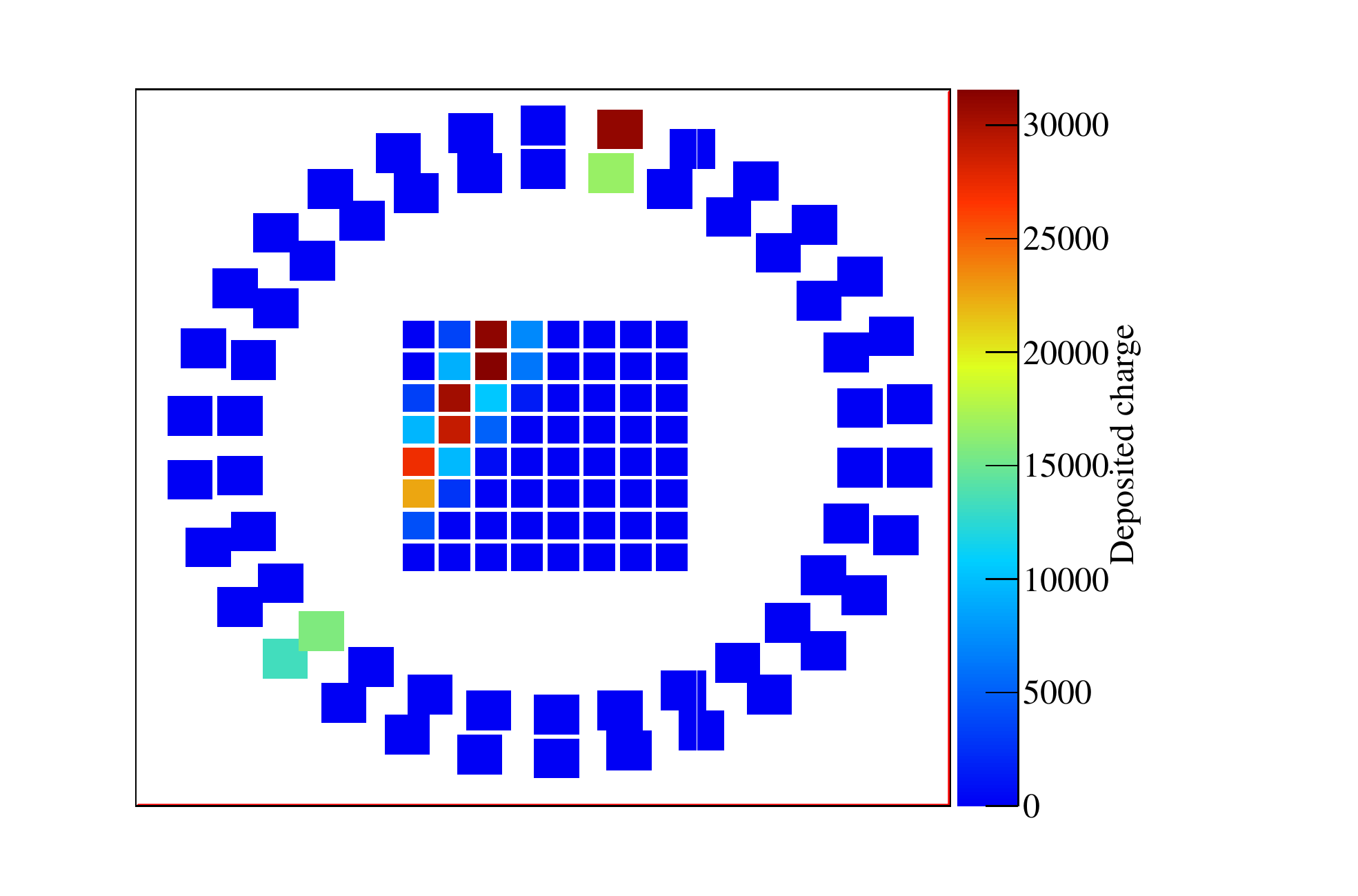}
\caption{(Left) Picture of the hodoscope in the beam line. (Right) Event display showing a projection of the hit position onto a plane perpendicular to the beam axis.}
\label{fig:hodoskop}       
\end{center}
\end{figure}

In order to be able to distinguish cosmic rays from \pbar\ annihilations, the central \Hbar\ detector is surrounded by a hodoscope consisting of 30 bars of plastic scintillator 5 mm thick, 10 mm wide and 120 mm long placed at $\sim 50$ mm radius around the beam axis parallel the beam directions, which are read out by one silicon photomultiplier on each side (cf. Fig.~\ref{fig:hodoskop} left). All 60 channels of the horoscope as well as the 64 channels of the central \Hbar\ detector are read out by FlashADCs (CAEN V1742) and the hit pattern is determined by software. Fig.~\ref{fig:hodoskop} (right) shows the hit pattern of both detectors projected onto a plane perpendicular to the beam direction. The color code corresponds to the energy deposit in each detector. The two rings represent the output of the downstreams and upstreams detectors of each hodoscope bar. The figure shows the track of a cosmic ray penetrating the hodoscope in a straight line. A \pbar\ annihilation would lead in more than 50\% of the events to the creation of three or more pions which should be clearly distinguishable from cosmic rays.

\section{Outlook}
\label{sec:3}

After a first test of the setup with the new particle detectors in summer 2012, a 1.5 year break in the antiproton production at CERN will take place. This time will be spent with analyzing the data and performing measurements with the apparatus using hydrogen. A monoatomic polarized hydrogen beam is currently being developed based on an RF discharge source \cite{Diermaier:2012fk}, a Q-mass spectrometer to detect the hydrogen atoms, and permanent sextupole magnets. Next attempts using an antihydrogen beam will commence in 2014.

\end{document}